\documentclass[10pt,aps,twocolumn,superscriptaddress]{revtex4-1}
\usepackage{amssymb,amsmath}
\usepackage{subfig}
\usepackage{graphicx}
\usepackage{color}
\usepackage[table,svgnames]{xcolor}
\usepackage[breaklinks]{hyperref}
\usepackage{listings}
\usepackage{tikz}
\usepackage{multirow}
\usepackage{lipsum}
\usepackage{booktabs}
\hypersetup{
    bookmarks=true,         
    unicode=false,          
    pdftoolbar=true,        
    pdfmenubar=true,        
    pdffitwindow=false,     
    pdfstartview={FitH},    
    pdfauthor={Kevin Multani},     
    colorlinks=true,       
    linkcolor=blue,          
    citecolor=red,        
}
\graphicspath{{figures/}}

\usepackage{csquotes}
\graphicspath{{files/}}

\tikzstyle{decision} = [top color=white, bottom color=blue!60,draw=blue!50!black!100, rectangle, text width=7em, text badly centered,  inner sep=3pt]

\tikzstyle{block} = [top color=red!40blue!10, bottom color =yellow!60, draw=yellow,rectangle, text width=8em, text centered, rounded corners, minimum height=4em]

\tikzstyle{longblock} = [top color=white, bottom color =yellow!60, draw=yellow!50!black!100,rectangle, text width=24em, text centered, rounded corners, inner sep=5pt]

\tikzstyle{bigblock} = [top color=white, bottom color =yellow!60, draw=yellow,rectangle, text width=18em, text centered, rounded corners, minimum height=2.5em,inner sep=0pt]

\tikzstyle{hugeblock} = [top color=white, bottom color =yellow!60, draw=yellow,rectangle, text width=20em, text centered, rounded corners, minimum height=0em, inner sep=5pt]

\tikzstyle{foco} = [top color=white, bottom color=red!50, draw=red!50!black,rectangle, text width=10em, text centered, rounded corners, minimum height=4em]

\tikzstyle{line} = [draw, -latex']

\tikzstyle{data} = [top color=red!10, bottom color= red!10, rectangle,  text width=2em, text badly centered, inner sep=0pt,minimum height=1.0em, inner sep=5pt]

\tikzstyle{data2} = [top color=green!10, bottom color=green!10, rectangle,  text width=6em, text badly centered, inner sep=0pt,minimum height=1.0em,inner sep =5pt]

\tikzstyle{data3} = [top color=blue!10, bottom color=blue!10, rectangle,  text width=6em, text badly centered, inner sep=0pt,minimum height=1.0em,inner sep =5pt]

\tikzstyle{longdata} = [top color=white, bottom color=red!50,draw=blue!50!black!100, rectangle, rounded corners, text width=25em, text badly centered, inner sep=2pt]

\tikzstyle{hugebox} = [top color=white, bottom color=red!50,draw=blue!50!black!100, rectangle, rounded corners, text width=34em, text badly centered, inner sep=2pt]

\newcommand{\eqdup}{\overset{\mathrm{D}}{=\joinrel=}}
\newcommand{\eqdde}{\overset{\mathrm{D+E}}{=\joinrel=}}

\begin{document}

\definecolor{dkgreen}{rgb}{0,0.6,0}
\definecolor{gray}{rgb}{0.5,0.5,0.5}
\definecolor{mauve}{rgb}{0.58,0,0.82}

\title{Automatic differentiation in quantum chemistry with an application to fully variational Hartree-Fock}
\author{Teresa Tamayo-Mendoza}
\affiliation{Department of Chemistry and Chemical Biology, Harvard University, 12 Oxford St, Cambridge, MA 02138, USA}

\author{Christoph Kreisbeck}
\email{christophkreisbeck@gmail.com}
\affiliation{Department of Chemistry and Chemical Biology, Harvard University, 12 Oxford St, Cambridge, MA 02138, USA}

\author{Roland Lindh}
\affiliation{Department of Chemistry-\AA ngstr\"{o}m, The Theoretical Chemistry Programme,}
\affiliation{Uppsala Center for Computational Chemistry, UC3, Uppsala University, Box 518, 751 20, Uppsala, Sweden}

\author{Al{\'a}n Aspuru-Guzik}
\email{aspuru@chemistry.harvard.edu}
\affiliation{Department of Chemistry and Chemical Biology, Harvard University, 12 Oxford St, Cambridge, MA 02138, USA}
\affiliation{Canadian Institute for Advanced Research, Toronto, Ontario M5G 1Z8, Canada}

\begin{abstract}
Automatic Differentiation (AD) is a powerful tool that allows calculating derivatives of implemented algorithms with respect to all of their parameters up to machine precision, without the need to explicitly add any additional functions. Thus, AD has great potential in quantum chemistry, where gradients are omnipresent but also difficult to obtain, and researchers typically spend a considerable amount of time finding suitable analytical forms when implementing derivatives. 
Here, we demonstrate that automatic differentiation can be used to compute gradients with respect to any parameter throughout a complete quantum chemistry method. We implement {\textit{DiffiQult}}, a fully autodifferentiable Hartree-Fock (HF) algorithm, which serves as a proof-of-concept that illustrates the capabilities of AD for quantum chemistry. We leverage the obtained gradients to optimize the parameters of one-particle basis sets in the context of the floating Gaussian framework.
\end{abstract}
\maketitle

\section{Introduction}

Automatic differentiation (AD) is a conceptually well-established tool to calculate numerical gradients up to machine precision \cite{Griewank2008-SIAM,Bischof2000}, by iteratively applying the chain rule and successively walking through the computational graph of numerically implemented functions. Therefore AD facilitates the computation of exact derivatives of coded algorithms with respect to all its variables without having to implement any other expression explicitly. AD circumvents challenges arising in traditional approaches. For example, the evaluation of analytical expressions tend to be inefficient and numerical gradients can be unstable due to truncations or round-off errors. AD has had a huge resurgence with the advent of deep learning \cite{Schmidhuber2015,Adams2015,Baydin2015Adi,Bengio2012}, and nowadays several libraries supporting AD are available \cite{Theano,tensorflow2015-whitepaper,Walther2012Gsw,Hascoet2013TTA,ADIFOR,Adams2015autograd}. Nevertheless, despite its advantages over conventional approaches, only a few examples exist where AD has been used outside the context of machine learning \cite{Jirari2016,Herard2013,Schuster2017,Cohen2017,Niemeyer2017}.

In this manuscript, we highlight the relevance of AD for quantum chemistry and demonstrate that AD provides a novel path forward to efficiently implement gradients for electronic structure methods. 
Gradients play a fundamental role in optimization procedures as well as for the computation of molecular response properties such as dipole moments polarizabilities, magnetizabilities, or force constants \cite{gauss2000}. 
Although, analytical expressions are available for some derivatives in many cases \cite{Almlof1985IJQC,Helgaker1988,Obara1986-JCP,Yukio2011}, researchers typically spend a considerable amount of time finding suitable analytical forms when implementing new quantum chemistry methods \cite{Baydin2015Adi}. 
Depending on the complexity of the electronic structure method this can result in a significant time delay between the development of the method and the implementation of its gradients. The implementation of geometrical gradients for MC-SCF \cite{kato197919,gauss2000} that has been published more than 10 years after the method itself \cite{veillard1967}.
Moreover, some analytical gradients can be too complicated to be handled manually. For example, the complex analytical derivatives of the FIC-CASPT2\cite{roos1990,roos1992} method, published more than two decades ago, has been only recently accessible through advances in automatic code generation\cite{Shiozaki2015,shiozaki2016}.

AD can marginalize the complexity of implementing gradients,
and thus can add significant value to quantum chemistry, as it allows to compute the gradients without coding them explicitly.
In the context of density functional theory, a combined approach using analytical expression together with AD has been successfully applied to compute higher order derivatives of the exchange-correlation term\cite{Ekstrom2010-JCTC}, as well as to compute response functions involving external magnetic fields and nuclear forces\cite{C0CP01647K}. Further, there are implementations of geometric derivatives on semi-empirical\cite{Steiger20051324} and quantum Monte Carlo methods\cite{Guidoni2012,Sorella2010}. To our knowledge, so far the applications of AD address only very specific aspects of quantum chemistry methods, while major components of the implementation still depend on analytical forms of the derivatives.

Here, we apply AD in a broader context and demonstrate that this technique is an efficient way to get arbitrary gradients for a complete quantum chemistry method with respect to any input parameter. To this end, we implement a fully autodifferentiable Hartree-Fock (HF) method which we distribute in our \textit{DiffiQult} software package\cite{DiffiQult}. We have selected HF since it is not only used in many electronic correlation methods as an initial step but contains complex mathematical operations and functions, such as calculating derivatives of eigenvectors or special functions. The latter is also relevant for more sophisticated quantum chemistry methods\cite{Jorgensen} and impose non-trivial requirements for suitable AD-libraries as they need to lay out the complete computational graph rather than calling black-box routines, for example, implemented in LAPACK \cite{lapack}. We illustrate the capabilities of \textit{DiffiQult} within the framework of a fully variational HF method, where we use a gradient-based optimization of the SCF-energy to optimize the parameters of the basis set within the Floating Gaussian framework\cite{Helgaker1988jcp,Frost1967}. Our implementation sets the basis for extending the \textit{DiffiQult} software package to include post-HF methods such as FCI and MP2, and to leverage higher order derivatives to obtain anharmonic corrections for rotational-vibrational spectroscopy. 

This paper is organized as follows:  In section~\ref{AD}, we provide a small review of the algebra behind automatic differentiation. In the section~\ref{FVHF} we introduce the fully variational Hartree-Fock method.
In section~\ref{Implementation}, we discuss in detail the key components to autodifferentiate the canonical Hartree-Fock algorithm, and 
explain how they were implemented in {\it DiffiQult}.
In Section~\ref{Results} we demonstrate the capabilities of our algorithm by optimizing the one-electron basis functions of small molecules.
Finally, in section~\ref{Conclusions}, we conclude with an outlook of future directions of {\it DiffiQult}, and a perspective of the role of AD in simplifying and accelerating the implementation of gradients of new quantum chemistry methods. 

\section{Automatic differentiation}\label{AD}
\begin{figure*}[t]
	            \begin{center}
	            \includegraphics[trim = {0.0cm 0.0cm 0cm 0.0cm},clip=True,scale=0.45]{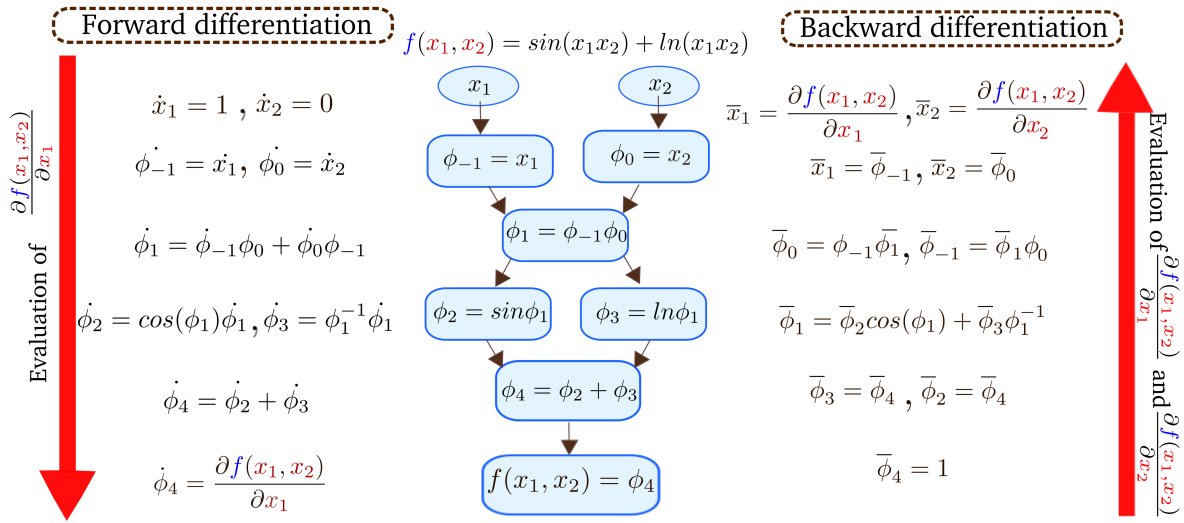}
	            \end{center}
            \caption{\small The center depicts the computational graph of a simple function $f(x_1,x_2)$ and its elementary operations. On the left and right, we illustrate the differentiation steps of forward and backward mode, respectively. In forward mode, the function evaluated at a given value ${\bf a}$, and the differentiation of the intermediate variables, ($\dot\phi = \frac{\partial \phi_{i+1}}{\partial \phi_i}$) are computed by iterating through the elementary operations of the computational graph. The forward direction is indicated by the left arrow. In backward mode, the function is evaluated first at a given value and later all adjoints, $\overline{\phi}_i = \overline{\phi}_{i+1} \frac{\partial \phi_{i+1}}{\partial \phi_i}$ are computed by iterating backwards through the computational graph, indicated by the right arrow. Notice, in this mode, the derivatives of the two independent variables are computed together, whereas in forward mode each partial derivative $\frac{\partial f}{\partial x_i}$  has to be evaluated separately. In this example, to compute the entire gradient, the number of operations in backward mode is smaller than in forward mode.}\label{fig:Back-For}
        \end{figure*}

The idea behind automatic differentiation is that
every algorithm, independent of its complexity, consists of a series of simple arithmetic 
operations that have known analytical forms, such as 
sums, multiplications, sines, cosines, or exponents.
The sequence of these elementary operations is represented by a computational graph, see for example Figure~\ref{fig:Back-For}.
In this form, it is possible to compute the gradients of
the outputs of a function with respect to its inputs by applying the chain rule and evaluating the analytical derivatives of all of these elementary operations in a given order. This implies that AD libraries can differentiate the entire algorithm 
not only mathematical expressions, written in an explicit form,
but also all the control functions such as recursions, loops, 
conditional statements, etc.
Therefore AD computes the exact derivatives of
any code with machine precision \cite{Griewank2008-SIAM}.
In practice, there are two main forms to compute derivatives using 
the chain rule with AD tools: forward and backward mode. 
We illustrate both modes in Fig.~\ref{fig:Back-For}.

Forward mode is conceptually the easiest way to compute gradients.
It is based on the calculation of the derivatives of the intermediate variables  $\phi_i$ with respect to the input variables ${x}_j$, $\dot{\phi}_i = \frac{\partial \phi_i}{\partial {x}_j}$ .
Forward mode evaluates sequentially each elementary operation following 
the computational graph and subsequently applying the chain rule.
In this way, the differentiation of all the intermediate variables
with respect to their parameters
are computed alongside with the function ${\bf y}({\bf x})$.
Both the function and derivative are 
evaluated at a particular value ${\bf a}$.
The algorithm starts from obtaining the partial derivative $\dot{\phi}_1$ of 
the first elementary operation $\phi_1$ with respect to
the parameter $x_j$, evaluated in $\bf{a}$.
Then it proceeds to compute the partial derivatives of the next elementary 
operations.
An example of forward differentiation of a simple function is illustrated on the left side in Figure \ref{fig:Back-For}.
Hence in forward mode, the differentiation goes from right side to
left side of the chain rule,
\begin{eqnarray}\label{eq:forward}
\dot{y}_i &=&
\left.
\frac{\partial {\color{blue}y_i}}{\partial {\color{red} x_j}} 
\right|_{{\bf a}}
= 
\left.
\frac{ \partial {\color{blue}y_i}}{\partial \phi_k} 
\left({\frac{\partial \phi_k}{\partial \phi_{k-1}}} 
\cdot\cdot\cdot
\left({\frac{\partial \phi_2}{\partial \phi_1}}
\left(
{\frac{\partial \phi_1}{\partial {\color{red}x_j}}} 
\right) \right) \right) 
\right|_{{\bf a}}
\nonumber \\
&=& 
\frac{ \partial {\color{blue}y_i}}{\partial \phi_k} 
\left(
{\frac{\partial \phi_k}{\partial \phi_{k-1}}} 
\cdot\cdot\cdot
\left(
{\frac{\partial \phi_2}{\partial \phi_1}}
\left(
\left.
{\dot{\phi}_1({\color{red}x_j})} 
\right|_{{\bf a}}
\right) 
\right) 
\right)  
\nonumber \\
&=&
\frac{ \partial {\color{blue}y_i}}{\partial \phi_k} 
\left(
{\frac{\partial \phi_k}{\partial \phi_{k-1}}} 
\cdot\cdot\cdot
\left(
\left.{\dot{\phi}_2({\phi_1})} \right|_{\phi_1({\bf a})}
\right) 
\right).
\end{eqnarray}
To compute the gradient of the function
we need to apply this procedure for each partial derivative with respect to each parameter individually. 
Since {[\bf the]} forward mode relies on the generation of the derivatives of the elementary operations with respect to each individual input parameter, it is particularly suited for functions with few independent variables.

The second form is backward mode, which relies on the computation
of the derivatives of the outputs
with respect to its intermediate variables, called ``adjoints" ($\overline{\phi}_i =\frac{\partial y}{\partial \phi_i}$).
With this definition, we can use the chain rule to obtain adjoints of the previous
elementary operations
\begin{eqnarray}
\overline{\phi}_i &=& \frac{\partial f(\phi_{i+1}(\phi_{i}))}{\partial \phi_{i+1}} \frac{\partial \phi_{i+1}}{\partial \phi_i} = \overline{\phi}_{i+1} \frac{\partial \phi_{i+1}}{\partial \phi_i}.
\end{eqnarray}
In contrast to forward mode, here partial derivatives are computed 
following the opposite direction through the computational 
graph, Fig.~\ref{fig:Back-For}. The algorithm starts with the computation of
the adjoint of the last elementary operation $\overline{\phi}_k$ of the function evaluation, which can be seen as a measure for 
the sensitivity of the function $\bf y=f(\bf x)$
with respect to the intermediate variable $\phi_k$. By iterating this process backwards through the elementary operations we obtain the derivative, which is equivalent of evaluating the
chain rule form left to right given by the following expression
\begin{eqnarray}\label{eq:barkward}
\overline{x}_j 
&=&  
\left.
\frac{\partial {\color{blue}y_i}}{\partial {\color{red} x_j}} 
\right|_{{\bf a}}
= 
\left.
\left(
\left(
\left(
\left(
\frac{ \partial {\color{blue}y_i}}{\partial \phi_k} 
\right) 
{\frac{\partial \phi_k}{\partial \phi_{k-1}}} 
\right) 
\cdot\cdot\cdot
{\frac{\partial \phi_2}{\partial \phi_1}}\right) 
{\frac{\partial \phi_1}{\partial {\color{red}x_j}}} 
\right) 
\right|_{{\bf a}}
\nonumber \\
&=& 
\left(
\left(
\left(
\left(
\left.
\overline{\color{blue}y}_i(\phi_k)
\right|_{{\phi_k(\bf a)}}
\right) 
{\frac{\partial \phi_k}{\partial \phi_{k-1}}}  \right) 
\cdot\cdot\cdot
{\frac{\partial \phi_2}{\partial \phi_1}}
\right)
{\frac{\partial \phi_1}{\partial {\color{red}x_j}}} 
\right) 
\nonumber \\
&=& 
\left(
\left(
\left(
\left.
\overline{\phi_k}(\phi_{k-1})
\right|_{{\phi_{k-1}(\bf a)}}
\right)
\cdot\cdot\cdot
{\frac{\partial \phi_2}{\partial \phi_1}}
\right)
{\frac{\partial \phi_1}{\partial {\color{red}x_j}}} 
\right),
\end{eqnarray}
where we define $\overline{y} = 1$.
The last step of backward mode, after 
calculating all the intermediate adjoints, results in the partial derivatives of the function with respect to all parameters. This is illustrated on the right side of Fig.~\ref{fig:Back-For},
where we get the entire gradient of $f(x_1,x_2)$ with a single backward propagation evaluation.
Note, that intermediate adjoints can be used at a later stage to obtain the partial derivatives with respect to both variables resulting in a reduction of operations compared to the forward differentiation. This characteristic generally makes backward mode more efficient for a small
number of dependent variables.
For more details about the general differences over both modes and
its implementation, the reader might refer to Ref.~\cite{Bischof2000}.

In practical applications, the choice of either differentiation mode depends on the specific
form of the computational graph as well as on performance considerations taking 
into account the available computational resources. 
For example, backward mode requires to store the complete sequence of
operations of the algorithm to compute the function, and either 
to store or to re-calculate the intermediate variables. On the other hand, this
constraint might be compensated by the capability of the backward mode to efficiently evaluate derivatives of 
many parameters with fewer operations than forward mode.
In our particular case, this analysis is done by reviewing the relevant mathematical operations of 
a canonical HF algorithm for an efficient autodifferentiable implementation.

\section{Fully variational Hartree-Fock}\label{FVHF}

Our goal is to develop an autodifferentiable 
quantum chemistry method. In this manuscript, as an illustrative, yet useful example, we apply AD to obtain HF energy gradients to optimize parameters of Gaussian basis functions.
This algorithm will enable a more compact wavefunction representation while maintaining a comparable level of accuracy in energy achieved by using larger basis sets. Thus this method will provide tailored molecular one-electron basis functions with greater flexibility compared to atomic-centered Gaussian expansions.

In the HF method, the wavefunction is constructed as a linear combination of one-electron basis functions, where we minimize the energy by finding the appropriate expansion
coefficients.
The selection of these basis functions is critical to accurately reproduce the behavior of a system\cite{MEST}.
The most popular form of these basis functions is the atomic-centered contracted Gaussian functions, because the 
computation of one- and two-electron integrals have
well established forms for their implementations\cite{Gill1990,Reine2012,Libint2,Obara1986-JCP,Obara1986-JCP}.
Each of these functions is defined by the set of following parameters:
(i) the exponents $\{ \alpha \}$ defining the Gaussian width, 
(ii) the contraction coefficients $\{c\}$ defining the linear expansion of the Gaussian representation of the atomic orbital (AO), 
(iii) the coordinates $ {\bf A}\ $ defining the center of the Gaussians, and
(iv) the angular momentum ${\bf a}$ defining the order of the polynomial factor defined
by the coordinate system (e.g. Cartesian or spherical harmonics)\cite{Schlegel1995}.

These one-electron basis functions 
are generally obtained by energy optimizations of single
atoms with a certain level of theory and are intended to mimic AO~\cite{Jensen2013}.
However, when AOs are used as a basis for molecules
they require some degree of flexibility to describe more complex behaviors such as polarization effects.
Therefore, one usually selects a relatively 
large number of basis functions 
with AOs of different exponents and sufficiently high angular momentum~\cite{MEST}.
The drawback of using a larger basis set is the increase in numerical complexity, imposing limitations on actual calculations.

To get a fully optimized wavefunction, we can minimize not only the expansion coefficients but all types of variational parameters of the Gaussian AOs mentioned above, 
such as nuclear coordinates, Gaussian centers, contraction coefficients and exponents. 
A fully variational approach may be useful to reduce the number of basis functions and to obtain a more compact representation of the wavefunction.
This could either improve the calculation of molecular properties~\cite{Helgaker1988jcp,MEST,Szabo} or yield a better reference state for certain higher levels of theory~\cite{Taylor88}.
A prominent example of a variational approach is the so-called Floating Gaussians~\cite{Frost1967,Huber1979,Huber1981,Huber1988} in which the centers of the AOs are optimized.
Some authors have partially optimized
parameters such as the Gaussian widths for 
the valence shell~\cite{Helgaker1988jcp} or included extra orbitals around the region of the chemical bond~\cite{Neisius1981,Neisius1982}.
Furthermore, wavefunction parameters have been optimized within a higher level of theory~\cite{Tachikawa2000,Tachikawa1999}, such as 
CASSCF and MP2~\cite{Shimizu2011}, or HF over ab-initio 
molecular dynamics~\cite{Brussel2014}. These methods have implemented gradient or Hessian-based optimization with analytical nuclear derivatives.

In this paper, we present a fully variational approach~\cite{Tachikawa1999}, where we optimize the molecular orbitals with a quasi-Newton method. Here, the gradients are calculated using a fully autodifferentiable HF implementation. By taking advantage of the AD techniques, we can compute numerically exact gradients with respect to any parameter without the need of implementing analytical gradients explicitly.
This provides the flexibility to either simultaneously optimize the Gaussian exponents and positions, or to optimize them sequentially.

\section{Implementation}\label{Implementation}
\begin{figure}[!ht]
	            \begin{center}
	            \includegraphics[trim = {3.1cm 0cm 3.1cm 0cm},clip=True,width = 8cm]{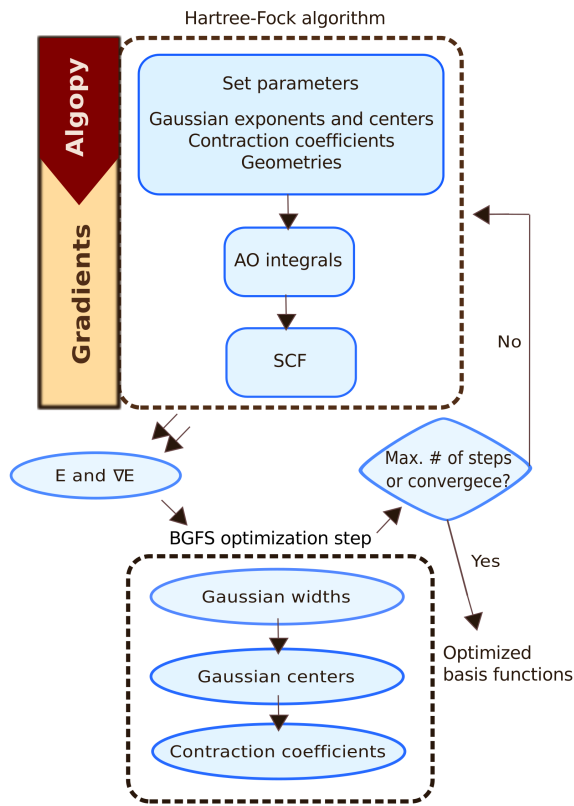}
	            \end{center}
            \caption{\small Diagram of the fully variational algorithm implemented in the {\it DiffiQult} package.}\label{fig:Algorithm}
        \end{figure} 
        
In the following section, we describe {\it DiffiQult}, 
a fully variational HF algorithm that calculates gradients employing AD.
In particular, we discuss options and constraints that need to be considered when applying AD in a quantum chemistry method. 
{\it DiffiQult} contains two main parts, (i) an autodifferentiable restricted HF implementation that provides
HF energies as well as its derivatives with respect to any input parameter and (ii) a gradient-based optimization of wavefunction parameters. The scheme of our fully variational Hartree-Fock implementation is
shown in Figure~\ref{fig:Algorithm}. 

The philosophy behind using AD is centered around the idea of saving human effort in the implementation of gradients. 
Ideally, we would like to choose a suitable AD library capable of getting gradients from AD just by adding
minimal changes to standard Python or C++ code.
This way, we would be able to significantly reduce the amount of time required for the implementation of gradients for new electronic structure methods. 
Moreover, we could link the AD library to existing electronic structure software packages for which analytical or numerical gradients might be inefficient. 
However, to what extent the aforementioned goals can be reached depends on the capabilities of the available AD libraries\cite{Theano,
tensorflow2015-whitepaper,Walther2012Gsw,Hascoet2013TTA,ADIFOR,Walter2011}. 
Each of the libraries differs in some aspects, but all of them have in common that they are restricted in some way, and it remains to be analyzed which ones match the requirements for a fully autodifferentiable quantum chemistry algorithm best. 
For example, the AD library needs to support linear algebra operations as well as calls to special functions such as the incomplete gamma function. Similarly, the AD modes, e.g. forward or backward mode, as discussed in section~\ref{AD} exhibit different properties, and one might be more suitable for a particular implementation than the other. 
A majority of the requirements are shared among various electronic structure algorithms, including HF.
Therefore the autodifferentiable HF implementation in {\it DiffiQult} serves as a proof-of-concept that helps us to develop an understanding of the capabilities of AD in quantum chemistry. 

In the following analysis, we review how the library and AD mode selection impact the implementation of {\it DiffiQult}.
Note that even operations that are simple in standard implementations,
such as control flow statements can raise challenges for AD. 
In HF, control statements are for example implemented to determine the required number of SCF steps for convergence. 
Depending on the input parameter, e.g., the molecular geometry, fewer or more iterations are needed to reach convergence, and thus the computational graph consists of fewer or more operations.
This has severe implications on the backward mode since for this mode the computational graph needs to be fixed before function evaluation.
We might circumvent some of these issues by hard-coding the number of steps
of the SCF, but this simple example demonstrates that the implementation of a fully autodifferentiable algorithm requires different considerations when compared to developing traditional software packages.

Another relevant aspect for the implementation of {\it DiffiQult} is the matrix operations. 
From the algorithmic point of view,
we should leverage vectorized operations whenever possible\cite{Giles2008}. 
Vectorized operations are taken as elementary operations in most
of the AD libraries,
and the computation of their derivatives are
typically stated in a simple vectorized form which avoids 
unnecessary storage and evaluations of intermediate variables. 
However in HF, as essentially in all wave-function based quantum chemistry methods,  
one and two electron integrals are typically defined by an element-wise array assignment
(eg. A[i,j]=k). This non-vectorized assignment represents a major challenge for backward mode regarding the amount of memory needed to store the computational graph as well as intermediate variables and derivatives.

The most critical component for our HF implementation in {\it DiffiQult} is matrix 
diagonalization. In some AD libraries such as {\it autograd}\cite{Adams2015autograd}, this matrix operation is considered as an elementary operation, which can be 
implicitly differentiated to obtain its adjoints and derivatives, respectively for each mode. In backward mode, the analytical expression for the adjoint of the eigenvectors is in principle available. However, the adjoints of the eigenvectors corresponding to repeated eigenvalues are not differentiable\cite{Walther2012Gsw,Giles2008}, see Appendix \ref{eigen}. Therefore, we cannot use backward differentiation for matrix diagonalization of systems with degenerate molecular orbitals. Since we aim to compute general molecular systems, we exclude backward mode for our implementation of {\it DiffiQult}. This leaves the forward mode as a possible option to circumvent the challenges of repeated eigenvalues.
For forward mode, the analytical expressions of the derivatives of eigenvectors are known,
even for the degenerate case\cite{Nelson1976}. 
This method relies on computing the $nth$-order derivative of the eigenvalues such that its diagonals are distinct and on computing the $n+1^{th}$-order derivative of the original matrix that needs to be diagonalized, see Appendix~\ref{eigen}. 
Therefore, we would need 
an implementation that would depend on a case-by-case analysis.
Alternatively,  Walter et al.\cite{Walter2012} proposed a general algorithm to compute the diagonalization, based on univariate Taylor polynomials (Appendix~\ref{utp})\cite{Griewank2008-SIAM,Walter2011}, implemented in Algopy\cite{Walter2011}. This approach is mathematically equivalent to forward differentiation and considers the repeated eigenvalue problem. 
For further details, we refer the reader to the Appendix. 

Even though we may consider reverse mode as a more efficient way for our implementation
given the large number of input parameters and the 
small number of output variables, after
this analysis, we conclude that our autodifferentiable HF implementation needs to 
be based on forward mode.
Regarding the AD library, we have chosen Algopy\cite{Walter2011}, since it supports both AD modes, provides matrix diagonalization for repeated eigenvalues, and requires only minimal modifications to autodifferentiate code implemented in plain Python.

Finally, once we have implemented an autodifferentiable HF algorithm in Algopy we use a
gradient-based optimization to optimize Gaussian centers, widths and contraction coefficients either together or separately. We use the Broyden\---Fletcher\---Goldfarb\---Shanno (BFGS) algorithm\cite{nocedal2006numerical} implemented
in the {\it scipy} module. As recommended in Ref.(\cite{Tachikawa1999}), we take the natural logarithm to optimize the Gaussian exponents. 

\section{Results}\label{Results}
\begin{table}[b!] 
        \centering
        \begin{tabular}{c c c c c c  } \toprule  \hline
        Molecule & \multicolumn{5}{@{}c@{}}{
                \begin{tabular}{@{}>{\centering\arraybackslash}m{1.4cm}
                                   >{\centering\arraybackslash}m{1.4cm}
                                   >{\centering\arraybackslash}m{1.4cm}
                                   >{\centering\arraybackslash}m{1.4cm}
                                   >{\centering\arraybackslash}m{1.4cm}
                                   >{\centering\arraybackslash}m{1.4cm}
                                   >{\centering\arraybackslash}m{1.4cm}
                                   >{\centering\arraybackslash}m{1.4cm}
                                   >{\centering\arraybackslash}m{1.4cm}}
                    \multicolumn{5}{@{}c@{}}{Basis} \\ 
                    \midrule\hline
        			\multicolumn{2}{@{}l@{}}{
                   \begin{tabular}{@{}>{\centering\arraybackslash}m{1.4cm}
                                   >{\centering\arraybackslash}m{1.4cm}}
                    \multicolumn{2}{@{}c@{}}{STO-2G} \\ 
                    \midrule\hline
                    None  & Opt. $coef$, $\alpha$ and $\bf A$   \\
                   \end{tabular}
                   }
                   &\multicolumn{2}{@{}l@{}}{
                   \begin{tabular}{@{}>{\centering\arraybackslash}m{1.4cm}
                                   >{\centering\arraybackslash}m{1.4cm}}
                    \multicolumn{2}{@{}c@{}}{STO-3G} \\ 
                    \midrule\hline
                         None & Opt. $coef$, $\alpha$ and $\bf A$   \\
                   \end{tabular}
                   } & Reference: 3-21G\\
                \end{tabular}
            }  \\ 
            \midrule \hline
        \showrowcolors
        HF & 
        \multicolumn{5}{@{}l@{}}{
        \begin{tabular}{@{}>{\centering\arraybackslash}m{1.4cm}
                                   >{\centering\arraybackslash}m{1.4cm}
                                   >{\centering\arraybackslash}m{1.4cm}
                                   >{\centering\arraybackslash}m{1.4cm}
                                   >{\centering\arraybackslash}m{1.4cm}}
       -95.60 & -97.03  & -98.57 & -99.38
       & -99.46   \\
      	\end{tabular}
        }\\
          H$_2$O & 
        \multicolumn{5}{@{}l@{}}{
        \begin{tabular}{@{}>{\centering\arraybackslash}m{1.4cm}
                                   >{\centering\arraybackslash}m{1.4cm}
                                   >{\centering\arraybackslash}m{1.4cm}
                                   >{\centering\arraybackslash}m{1.4cm}
                                   >{\centering\arraybackslash}m{1.4cm}}
        
       -72.74 & -73.82&
       -74.96    &  -75.52 &  
       -75.59  
        \\            
      	\end{tabular}
        }\\
          NH$_3$& 
        \multicolumn{5}{@{}l@{}}{
        \begin{tabular}{@{}>{\centering\arraybackslash}m{1.4cm}
                                   >{\centering\arraybackslash}m{1.4cm}
                                   >{\centering\arraybackslash}m{1.4cm}
                                   >{\centering\arraybackslash}m{1.4cm}
                                   >{\centering\arraybackslash}m{1.4cm}}
        
       -53.82  &   -54.65  & 
       -55.45  & -55.83 &
       -55.87  \\            
      	\end{tabular}
        }\\
          CH$_4$ & 
         \multicolumn{5}{@{}l@{}}{
        \begin{tabular}{@{}>{\centering\arraybackslash}m{1.4cm}
                                   >{\centering\arraybackslash}m{1.4cm}
                                   >{\centering\arraybackslash}m{1.4cm}
                                   >{\centering\arraybackslash}m{1.4cm}
                                   >{\centering\arraybackslash}m{1.4cm}}
       -38.59 & -39.32
       & -39.72 & -39.96 &
       -39.98  \\                               
      	\end{tabular}
        }\\
         CH$_3$F & 
         \multicolumn{5}{@{}l@{}}{
        \begin{tabular}{@{}>{\centering\arraybackslash}m{1.4cm}
                                   >{\centering\arraybackslash}m{1.4cm}
                                   >{\centering\arraybackslash}m{1.4cm}
                                   >{\centering\arraybackslash}m{1.4cm}
                                   >{\centering\arraybackslash}m{1.4cm}}
        
       -133.09 & -134.96  &
       -137.17 & -137.43  &
       -138.28  \\                               
      	\end{tabular}
        }\\
         CH$_2$O & 
         \multicolumn{5}{@{}l@{}}{
        \begin{tabular}{@{}>{\centering\arraybackslash}m{1.4cm}
                                   >{\centering\arraybackslash}m{1.4cm}
                                   >{\centering\arraybackslash}m{1.4cm}
                                   >{\centering\arraybackslash}m{1.4cm}
                                   >{\centering\arraybackslash}m{1.4cm}}
      -109.02 &  -110.54  &
      -112.35 & -112.72 &
       -113.22 \\                               
      	\end{tabular}
        }\\
        \bottomrule
 \hline
      \end{tabular} \caption{\small Hartree-Fock energies for test molecules of the non-optimized and optimized (sheme~(B)) basis sets STO-2G and STO-3G. As a reference, we show results for the larger 3-21G basis set.}\label{results}
    \end{table}

\begin{figure}
\centering 
\includegraphics[width = 0.5\textwidth]{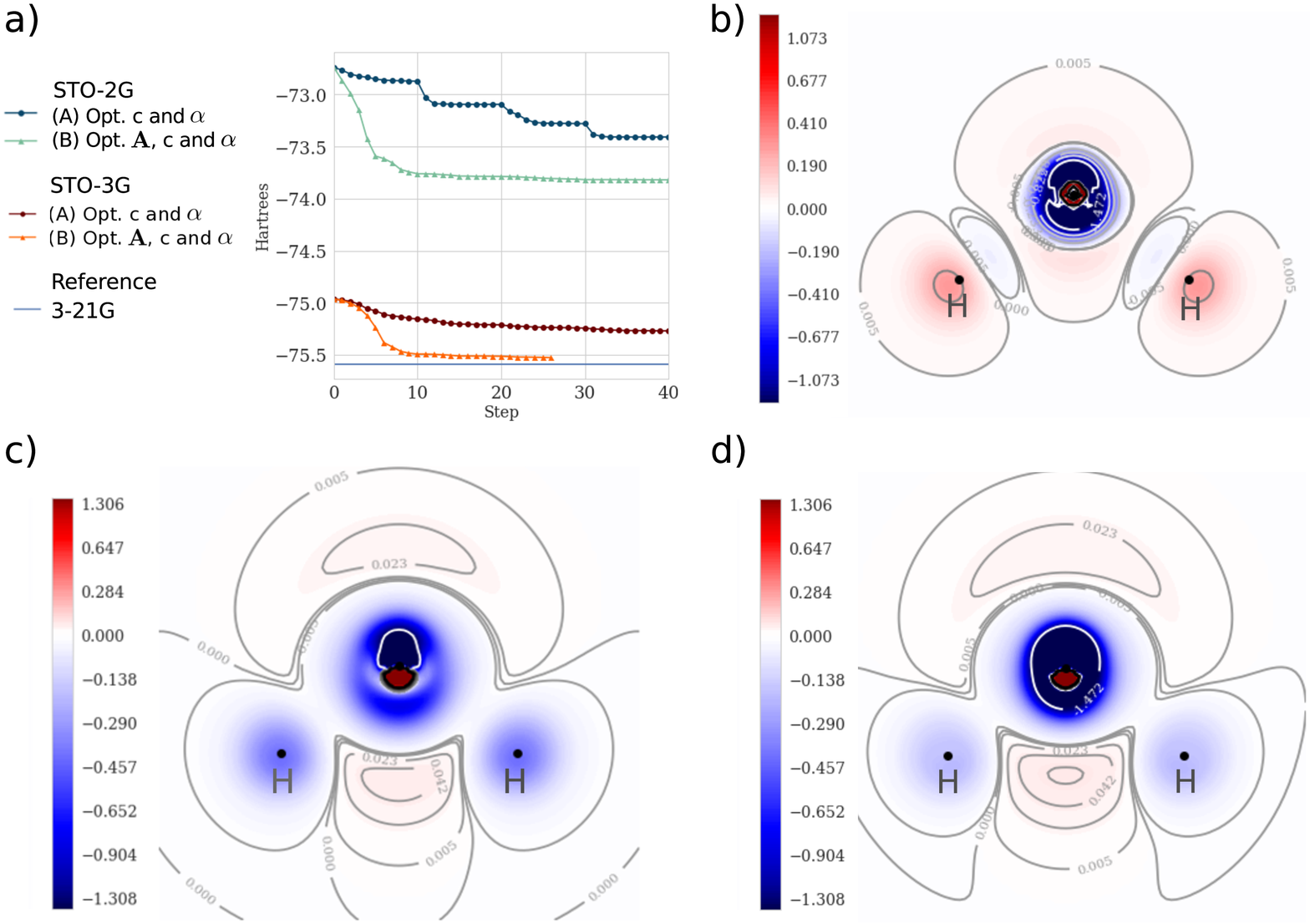}
\caption{\small One-electron basis function optimizations of H$_2$O a) Optimization steps of STO-3G and STO-2G with different optimization schemes. (A) Optimizing exponents and coefficients 10 steps each twice, and (B) optimizing exponents and coefficients together, followed by optimizing positions. 
b) contour of the difference in electronic density between the STO-3G basis and the optimized one-electron basis functions under scheme (B). c) contours of the difference in electronic density between the conventional STO-3G basis set and the reference 3-21G basis set. d) contour of the difference in electronic density between the optimized (scheme~(B)) STO-3G basis set and the reference 3-21G basis set. An increasing of density on the optimization is displayed in blue and a decreasing in red.
}\label{fig:Optimizedpaths}
\end{figure}

We tested our implementation by optimizing the STO-2G as well as the STO-3G minimal basis sets for the  small molecules H$_2$O, HF, NH$_3$, CH$_4$, and CH$_2$O. Since we can obtain the gradients with respect to any input parameter, 
{\it DiffiQult} has the freedom to select different optimization procedures.
Here, we illustrate two schemes: 
(A) optimizing the Gaussian exponents and contraction coefficients, 
sequentially, and
(B) optimizing the contraction coefficients and exponents, 
followed by an optimization of the 
Gaussian centers.
For the example of H$_2$O, the improvement of the energies for each optimization step are illustrated in Figure~\ref{fig:Optimizedpaths}.
We find that the most efficient way to optimize the HF-energy is by employing scheme~(B) since it already converges after ten basis set optimization steps. 
For a general assessment of
which method works better than the other, we need to consider the
trade-off between reaching fast convergence and computation time.  For example,
scheme (B) requires more time to perform a single optimization step since it requires to compute more gradients for a larger number of parameters due to the line-search procedure within the BFGS algorithm. 
All optimizations were done until finding an infinite norm of the gradient of 10$^{-5}$ or a total of 10 steps respectively.
Table \ref{results} displays HF-energies for optimized and non-optimized basis sets for selected small molecules. 
The optimization scheme~(B) results in an improvement of up to 0.18~Hartrees per electron.

The minimal basis sets we optimized
lacked in flexibility to correctly represent polarization and dispersion in the core molecular orbitals.
For instance, hydrogen contains only a single atom centered s-type orbital that is insufficient to display changes in the density induced by the surrounding charges.
By optimizing the parameters of the atomic orbital of each hydrogen we can partially take into account polarization effects.
As is depicted in Fig.~\ref{fig:Optimizedpaths}~(b), the optimization of the basis STO-3G of H$_2$O shifts the electronic density from the regions around the hydrogen atoms towards the bond regions. 
In the case of the oxygen, as it is shown in Fig.~\ref{fig:Optimizedpaths}~(c) and~(d), the number of basis function is not sufficient 
to represent both the bond and the density around the oxygen itself, even after the optimization.
Therefore, corrections to the core atomic orbitals of oxygen 
would need to include a greater number and higher angular momentum one-electron functions.

Finally, we discuss a common convergence problem of the HF method that can affect our fully variational HF implementation. 
It appears in our examples mainly in the line-search, 
when the optimizer by chance
tests parameter for which the SCF is difficult to get converged. In principle, this problem can be circumvented by suitable convergence strategies in the HF algorithm.

\section{Conclusion and Perspective}\label{Conclusions}
 
AD offers a promising alternative solution for computing accurate derivatives of electronic structure methods. 
In this manuscript, we implemented {\it DiffiQult}, which serves as a proof-of-concept that helps us to develop an understanding of the capabilities of AD in quantum chemistry. Specifically, we presented and discussed the use of AD in the context of a fully variational HF method. By using the \textit{Algopy} AD-library, we implemented gradients with just minimal adjustments in the source code of the canonical HF-method. With these gradients at hand, we are able to fully minimize the energy with respect to any parameter of the Gaussian one-particle basis. As a result, we capture, to some extent, polarization effects with a reduced number of atomic orbitals. Since the essential functions of many quantum chemistry methods are similar to the ones present in HF, we plan to extend {\it DiffiQult} to post-HF methods in a future work. 

\textit{DiffiQult} can be seen as a general tool to obtain molecular tailored basis functions, that can be used as a starting point for other variational methods, e.g. FCI. 
An emerging application could be in the
field of quantum computing for quantum chemistry\cite{Peruzzo}, where the size of one-electron basis function is constrained by the number of qubits available in state-of-the-art hardware\cite{Barends2014,Barends2015,Riste2015,Corcoles2015,Malley2016}. Thus, experimental demonstrations of quantum algorithms for chemistry have been limited to conventional minimal basis sets \cite{Malley2016,Kandala}. Here, the fully variational setting of \textit{DiffiQult} could offer the advantage to optimize initial parameters of atomic orbitals, which could increase the accuracy of variational quantum chemistry simulations\cite{Peruzzo}, while keeping the number of basis functions small.

Given the flexibility of AD libraries, the current approach to use AD in chemistry could be extended in many ways. For example, in general, it is possible to mix different differentiation modes and approaches. We could use forward mode to compute gradients of element-wise matrix definitions (or in general for complex computational graphs), and use backward mode for vectorized functions. Furthermore, AD could be combined with symbolic algebra and automated code generation \cite{Hirata2004jcp-12197,Jasen1991Th-Ch-Acta,TCE} build a general tool-chain to create autodifferentiable software packages, which could potentially decrease the human time spent on code development by orders of magnitude.

\section{Acknowledgments}
This project was supported by the NSF award number 1464862. 
R. L. acknowledges support by the Swedish Research 
Council (grant 2016-03398).
T. T-M. thanks CONACyT for scholarship No. 433469.
T. T-M. appreciates the rich discussions with Jhonathan Romero Fontalvo and Benjamin Sanchez Langeling.

\bibliography{paper}

\appendix
\subsection{Derivatives of eigenvectors}\label{eigen}
The differentiation
of the matrix diagonalization
$\bf{D^T A D = \Lambda}$ is implemented in several libraries
as an elementary operation.
In the following derivations, we provide a broad outline of the methods and their limitations in differentiating matrix diagonalization for forward and backward modes. For further detail, we refer to Refs.~\cite{Giles2008,Walter2012,Walter2012Structured,Nelson1976}.

In the case of backward mode, the adjoint of eigenvectors $\overline{\bf{A}}$ can be obtained by implicitly 
differentiating the eigenvalue problem, resulting in the following expression
\begin{eqnarray}
\overline{\bf{A}} &=&
\left( {\bf D}^{-1} \right)^T
\left(\overline{{\bf \Lambda}}
 +
{\bf F} \circ 
{\bf D}^{T}  {\bf D} \right){\bf D}^T,
\end{eqnarray}
where $F$ is zero along the diagonal and $F_{ij} =  (\lambda_j - \lambda_i)^{-1}$ for $i\neq j$.
Note that $F_{ij}$ diverges for repeated eigenvalues.
Therefore, we cannot use backward differentiation for systems with degenerate molecular orbitals.

In the case of forward mode, one method to compute the derivatives $\bf \dot{D}$ is by finding 
the appropriate matrix ${\bf C}$, such that,
\begin{equation}
{\bf \dot{D}} = {\bf DC}.
\end{equation} 
For non-degenerate eigenvalues, ${\bf C}$ can be obtained by
\begin{eqnarray}
c_{ij} &=& \frac{{\bf D}^T_i {\dot A} {\bf D}_j}{2(\lambda_i- \lambda_j)} \phantom {m}  i \neq j \textnormal{ and }
c_{ii} = -\frac{1.0}{D_{ii}} \sum^n_{m} D_{im}c_{mi}   \phantom {i \neq j}.
\end{eqnarray}
It is possible to extend this approach to the degenerate case though this requires calculations of higher order derivatives
\begin{eqnarray}
c_{ij} = \frac{{\bf D}^T_i { A ''} {\bf D}_j}{2(\dot{\lambda_{ii}}- \dot{\lambda_{jj}})}.
\end{eqnarray}
In cases in which derivatives of eigenvectors are repeated, 
there is a similar expression that again requires computing
derivatives of the next order.
This procedure needs to be repeated up to the point in which all diagonal terms on the $n$th order derivatives of eigenvalues are distinct.
As a result, such implementation will require computations of higher-order derivatives as well as a case-by-case analysis depending on the problem at hand.

An alternative algorithm was proposed by Walter et al.\cite{Walter2012} which has been implemented in the 
the Python library \textit{Algopy}\cite{Walther2012Gsw}.
This library utilizes the
univariate Taylor polynomial arithmetic\cite{Griewank2008-SIAM,Bucker2005ADA}
to compute higher-order derivatives and applies a general algorithm for the eigenvalue
problem. 
This formalism offers a scheme to compute higher order derivatives of explicit and implicit functions.
For each dependent and independent variable, there is a Taylor polynomial whose coefficients correspond to the value of the variable and its derivatives.
With these, we can construct a system of equations that contains different differentiation order to compute the coefficients of the polynomial corresponding to the dependent variable.
The set of equations is solved by using the the Newton-Hensel lifting method, also called Newton's method\cite{Griewank2008-SIAM}.
In our specific case, to cover the degenerate case, 
these set of equations are computed systematically
by blocks of $D$ and $\Lambda$ of the same eigenvalue or derivative.

\subsection{Univariate Taylor Arithmetic}\label{utp}
Forward differentiation can be formulated 
with the Taylor ring arithmetic\cite{Griewank2008-SIAM}. 
This arithmetic allows us 
to implicitly differentiate a function and compute higher order derivatives.
In the following, consider a function $y(t) = F(x(t))$ where
$F: \mathbb{R}^{n} \longmapsto \mathbb{R}^{m}$, for a given smooth curve
\begin{eqnarray}
x(t) = \displaystyle\sum^{D-1}_{d=0}  x_d t^d
\end{eqnarray}
with $t \in (-\epsilon,\epsilon)$.
Using Taylor's
theorem we obtain
\begin{equation}
y(t) = \displaystyle\sum^{D-1}_{d=0} y_d t^d
+ \mathcal{O}(t^D)
\end{equation}
where
\begin{equation}
y_d := \frac{1}{d} \frac{\partial^d}
{\partial t^d} 
\left. x(t) \right|_{t=0}  
\end{equation}
In this form the coefficients $\mathbb{R}^{D \times n}$ included in $x(t)$
are mapped to the coefficients $\mathbb{R}^{D \times m}$  yielding a approximate expression for $y(t)$.
Note that these sets of polynomials defined by
\begin{equation}
\mathcal P_d = \left \{ \left. x(t) = \displaystyle\sum_{j=0}^{d-1} x_j t^j \right| x_j \in \mathbb{R} \right\}
\end{equation}
form a commutative ring for every order $d > 0$.
This allows us to obtain Taylor polynomials from the binary operations of addition, multiplications and subtraction operations.
Furthermore, we can apply the rules of multivariate calculus to any continuous and d-time differentiable function $F$.
These functions and their derivatives can be mapped to their corresponding Taylor coefficients with
the extension function $E_D(F)$,
$E_D(F): \mathcal{P}_D^n \longmapsto \mathcal{P}_D^m$,
defined by,
\begin{eqnarray}
[y]_D &=& E_D(F)([x])_D = \displaystyle\sum_{d=0}^{D-1} y_d t^d  \nonumber \\
 &=& \displaystyle\sum^{D-1}_{d=0} \frac{1}{d!} \frac{\partial^d}{\partial d ^t}_d 
 \left. F \left( \displaystyle\sum^{D-1}_{d=0}  x_d t^d \right) \right|_{t=0}  T^d
\end{eqnarray}
with $[x]_D \equiv [x_0,x_1,...,x_{D-1}]$ and 
$[x]_{d:D} \equiv [x_d,x_1,...,x_{D-1}]$.
Furthermore, this extended function follows the rules of composite functions, 
$F(x) = (H \circ G)(x) $ such that,
\begin{equation*}
E_D(F) = E_D(H) \circ E_D(G).
\end{equation*}
This relation sets the foundation for forward differentiation that allows us to
use the chain rule with a set of elementary operations.
Furthermore, we can extend this formalism to implicit equations  $0= F(x,y) \in \mathbb{R}^M$.
For a given Taylor polynomial $[x]_D$, we can find the polynomial $[y]_D$ which is defined by a
system of equations, up to order $D$
\begin{equation}
0\eqdup E_D(F)([x]_D,[y]_D),
\end{equation}
where $ [x] \eqdup [y]$ if $x_d = y_d$  for $d=0,...,D-1$.
Once we know the coefficients $[y]_D$,
we can solve the next $E$ orders where $1 \leq E\leq D$, using 
the Newton-Hensel lifting method\cite{Griewank2008-SIAM}. 
The method solves the next set of equations based on the previous
orders, such that
\begin{eqnarray}\label{sys}
0\eqdde E_{D+E}(F)([x]_{D+E},[y]_{D+E}).
\end{eqnarray}
If the function F is differentiable and 
$F_y (x_0,y_0)$ is invertible, then the new coefficients are calculated
by $ {[ \Delta y]}_{E} = [\Delta y]_{D+E} - [\Delta y]_{E}$,
which can be computed by the expression
\begin{eqnarray}
{[ \Delta y]}_{E} = [F_y]_E^{-1} [\Delta F]_E.
\end{eqnarray}
where $E_{D+E}(F)([x]_{D+E},[y]_{D+E}) \eqdde [\Delta F]_{E}^{-1} T^D$ and
$[F_y]_E = E_E \left({\frac{\partial F}{\partial y}} \right) ([x]_{E},[y]_{E})$.
These equations allow us to employ the Newton-Hensel method to 
compute the Taylor coefficients of f$[y]$.
This method solves iteratively the system of equations eq.~(\ref{sys}), by either doubling the
number of coefficients following Newton's method, or
solving the coefficient once at a time, using sequentially the lifting Newton-Hensel algorithm. 

\end{document}